\documentclass[a4paper,12pt,usenatbib]{article}
\usepackage{setspace}
\usepackage[dvipdfmx]{graphicx}
\usepackage{amsmath,amssymb}
\usepackage{mathrsfs}
\usepackage{mathptmx}
\usepackage{mathtools}
\usepackage{bm}
\usepackage{hhline}
\usepackage{subfigure}
\usepackage{threeparttable}
\usepackage{txfonts}
\usepackage{indentfirst}
\usepackage[tablesfirst]{endfloat}
\usepackage{url}
\usepackage{here}
\usepackage{wrapfig}
\usepackage{comment}
\usepackage{multirow}
\usepackage{tcolorbox}
\usepackage{colortbl}
\tcbuselibrary{breakable, skins, theorems}
\usepackage[sectionbib,sort&compress]{natbib}
\setcitestyle{aysep={,},citesep={;}}
\usepackage{comment}
\usepackage{algorithm}
\usepackage{algpseudocode}
\usepackage[top=25.4truemm,bottom=25.4truemm,left=25.4truemm,right=25.4truemm]{geometry}
\begin{document}
\begin{center}
{\rm {\LARGE {\bf Exact sequential single-arm trial design \\ with curtailment for binary endpoint}}}
\end{center}
\begin{center}{
\text{TASUKU INAO $^\ast$} 
}\end{center}
\textit{Department of Biostatistics, Graduate School of Medicine, Hokkaido University, Sapporo, Japan} \\
\text{E-mail: inao.tasuku.z3@elms.hokudai.ac.jp} \\
\begin{center}{
\text{ISAO YOKOTA}
}\end{center}
\textit{Department of Biostatistics, Graduate School of Medicine, Hokkaido University, N15, W7, Kita-ku, Sapporo, Hokkaido, Japan}
\\
\text{E-mail: yokotai@pop.med.hokudai.ac.jp}

\vspace{4truemm}

\noindent \hrulefill

\noindent { \bf SUMMARY}
\vspace{5mm}
\begin{spacing}{2.0}
For trials for rare cancer, the single-arm design with binary response is preferred incorporating sequential monitoring. It tends to be hard to conceal the number of responders completely, and it may be appropriate to present the prespecified thresholds and to monitor each patient to determine if the trial is a success or not. In this paper, we propose the statistical monitoring method with the threshold of responders for efficacy stopping fixed over the number of participants. To maintain the error rate, it is exactly calculated using the negative binomial distribution. The bias-adjusted point estimator is also proposed. The simulation experiments showed that the maximum sample size of our exact sequential design was less than an exact fixed design, and the average sample number of our design was as small as conventional Simon's two-stage design or spending function approach. Also, we observed a small bias in proposed estimator and the convenient methods of confidence interval.
\end{spacing}

\noindent \hrulefill

\noindent { \bf KEYWORDS}

\noindent \hspace{1mm} deterministic curtailment, exact test, maximum sample size, sequential monitoring

\newpage
\section{Introduction}
\label{introduction}
\begin{spacing}{2.0}
When we assess the efficacy of the new regimen in the phase II single-arm oncology clinical trial, especially rare or childhood cancer, the primary endpoint is set to be tumor response as a binary outcome. To conduct the trial more efficiently, we often adapt the interim monitoring. Most popular one is Simon’s two-stage design (\citealp{simon1989optimal}; \citealp{ivanova2016nine}), which enables early stopping for futility when the number of response is less than prespecified cut-off number.
The timing and thresholds of interim and final analysis can be derived by the minimax or optimal criteria, which are calculated to minimize the maximum sample size and the average sample number, respectively, under the null hypothesis. To further consider more complicated situations, the extended designs derived from Simon's 2-stage design have also been developed (\citealp{jones2007adaptive}; \citealp{kunz2012curtailment}).
Simon's design was extended to allow early stopping for efficacy (\citealp{chen2008optimal}; \citealp{MANDER2010572}; \citealp{https://doi.org/10.1002/pst.1965}).
On the other hand, the phase III trial often used Lan-DeMets’ $\alpha$-spending function method for an efficacy stopping (\citealp{gordon1983discrete}). 
The spending function approach can be applied for one-sample binary test using asymptotic test statistics (\citealp{jennison1999group}).

The outcome, such as tumor response, tends to be unblinded and found out by patients.
Thus, the trialists may find out that the number of responders meets its threshold if it exceeds the far expectation.
Rather than concealing the number of responders, it may be appropriate to present the pre-specified thresholds and to monitor each patient to determine if the trial is a success or not.
To realize such a design, conventional sequential monitoring methods have two practical difficulties.
First, the threshold of responders depends on the number of trial participants.
Under staggered entries, there are still subjects who have already enrolled and who have not yet been known the response status.
If subjects who are under follow-up are considered as non-responder and analyzed, statistical power will be low.
Otherwise, if these subjects are omitted on the denominator, the decision of efficacy stopping may be overturned after revealing the response status for all participants.
Second, when using the asymptotic sequential method or asymptotic test statistics, the actual error rate $\alpha$ may exceed the nominal level due to finite small samples
 (\citealp{fleming1982one}; \citealp{a2001sample}).

In this paper, we propose the statistical monitoring method with the threshold of responders for efficacy stopping fixed over the number of participants. To maintain the error rate, it is exactly calculated using the negative binomial distribution.
It will be apparent how many additional responders are needed for the study to be successful.
Also we impose the threshold of futility stopping using deterministic curtailment.
That is, if the number of responders does not reach the threshold even though all of the remaining participants have responded, the trial is stopped due to futility.
As making the threshold being fixed over the trial, it is expected to be a comprehensive monitoring for trial sponsors.

\end{spacing}

\section{A framework of proposed design}
\label{design}
\begin{spacing}{2.0}
Set the statistical hypotheses as follows: $\mathcal{H}_0:p=p_0,\mathcal{H}_1:p=p_1$, where $p_0$ and $p_1$ are the null and expected response rates. At the $k (k=1,2,\cdots,K)$th stage of interim monitoring, let $S_k$ be the number of responders, respectively. The threshold of the responders $u$ to stop for efficacy is constant throughout the monitoring.
If $S_k\geq u$ in the $k$th stage, stop the monitoring and reject $\mathcal{H}_0$.
Suppose that an interim analysis is performed on each patient's observation of response status. That is, the number of subjects in the $k$th interim analysis is identical to $k$.
In the earlier period of interim monitoring, in which the number of participants is less than the threshold of responders, $k<u$, the interim analysis is not needed because stopping for efficacy never occurs.
When the interim analysis is skipped after $k \ge u$, the overshooting of responders may occur, but the decision to prove efficacy or not is identical.
The stopping probability in the $k$th interim monitoring can be calculated exactly based on the following negative binomial distribution due to its data generating process:
\begin{equation}
\label{eq:1}
NB(S_k,k;p)=\binom{k-1}{S_k-1} p^{S_k}(1-p)^{k-S_k}.
\end{equation}
The actual $\alpha$ error rate and power can be easily calculated using the cumulative distribution function of (\ref{eq:1}):
\begin{align*}
    \alpha =& \sum_{k=u}^K NB(u,k;p_0)\\
    1-\beta =& \sum_{k=u}^K NB(u,k;p_1).
\end{align*}
For example, when we evaluate $(u,K)=(3,4)$ under the situation $(p_0,p_1)=(0.1,0.55)$, the $\alpha$ error rate and power is calculated as follows: 
\begin{equation*}
\begin{split}
\alpha&=NB(3,3;p_0)+NB(3,4;p_0)=0.0037 \\
1-\beta&=NB(3,3;p_1)+NB(3,4;p_1)=0.3909
\end{split}
\end{equation*}

The threshold $u$ and the maximum sample size $K$ are set by grid search procedure referring to the nominal levels of $\alpha$ and $\beta$.
To minimize the sample size, start the search with a small $u$.
Under fixed $u$, increase $K$ and calculate $\alpha$ and $\beta$ error rate.
Find the pair $(u,K)$ satisfying that both $\alpha$ and $\beta$ error rate are below each nominal level.
If no pairs are satisfying that condition, search for one large $u$.
If we find the pairs satisfying that condition, the sample size may be set as the minimum of $K$.
Overshooting of enrollment is allowed to the range of the largest $K$.

This procedure is illustrated in the example with the nominal one-sided $\alpha$ and $\beta$ levels are 0.025 and 0.2, respectively, under the situation $(p_0,p_1)=(0.1,0.55)$.
Table \ref{tab;example_0.1_0.55_abeta} shows the values actual $\alpha$ error rate $\alpha$ and power $1 - \beta$ for each pair of $u$ and $K$.
First, for $u=2$, only $K=2$ satisfied the nominal $\alpha$ level, but the power is only 0.3.
Next, fixed at $u=3$, $K=3,4,5,6$ satisfied the nominal $\alpha$ level.
Nevertheless, no appropriate pair was found, as the power is still a maximum of 0.74 at $K=6$.
Thus, searching at $u=4$, we could find that $K=4,\cdots,11$ satisfies the nominal $\alpha$ level and $K=9,10,11$ have the power over 0.8.
Therefore, the threshold for efficacy stopping and maximum sample size are set at 4 and 9.

\begin{table}[htbp]
\centering
  \rotatebox{90}{
    \begin{minipage}{\textheight}
\caption{$\alpha$ error rate and power in proposed method with $(p_0,p_1)=(0.1,0.55)$.}
\begin{threeparttable}
\begin{tabular}{c|c|ll|c|ll|c|ll} \hline \hline
 &  & \multicolumn{8}{c}{$u$} \\ \cline{3-10}
 &  & \multicolumn{2}{c}{2}& &\multicolumn{2}{c}{3}& &\multicolumn{2}{c}{4} \\ \hline \hline
\multirow{11}{*}{$K$}
& 2 & $\alpha=0.01$ & \cellcolor[gray]{0.8}$1-\beta=0.3$ & & & & \\ \hhline{|~|-|-|-|-|-|-|-|-|-|}
& 3 & \cellcolor[gray]{0.8}$\alpha=0.028$ & \cellcolor[gray]{0.8}$1-\beta=\times$ & &$\alpha=0.001$ & \cellcolor[gray]{0.8}$1-\beta=0.16$ & & &  \\ \hhline{|~|-|-|-|-|-|-|-|-|-|}
& 4 & \cellcolor[gray]{0.8}$\alpha=0.05$ & \cellcolor[gray]{0.8}$1-\beta=\times$& &$\alpha=0.004$ &\cellcolor[gray]{0.8}$1-\beta=0.39$& &$\alpha<0.001$ & \cellcolor[gray]{0.8}$1-\beta=0.09$ \\ \hhline{|~|-|-|-|-|-|-|-|-|-|}
& 5 & \cellcolor[gray]{0.8}$\alpha=0.08$ & \cellcolor[gray]{0.8}$1-\beta=\times$& &$\alpha=0.009$ &\cellcolor[gray]{0.8}$1-\beta=0.59$& &$\alpha<0.001$ & \cellcolor[gray]{0.8}$1-\beta=0.25$ \\ \hhline{|~|-|-|-|-|-|-|-|-|-|}
& 6 & \cellcolor[gray]{0.8}$\alpha=0.11$ & \cellcolor[gray]{0.8}$1-\beta=\times$& &$\alpha=0.016$ &\cellcolor[gray]{0.8}$1-\beta=0.74$&&$\alpha=0.001$ &\cellcolor[gray]{0.8}$1-\beta=0.44$ \\ \hhline{|~|-|-|-|-|-|-|-|-|-|}
& 7 & \cellcolor[gray]{0.8}$\alpha=0.15$ & \cellcolor[gray]{0.8}$1-\beta=\times$& &\cellcolor[gray]{0.8}$\alpha=0.0256$ &\cellcolor[gray]{0.8}$1-\beta=\times$& &$\alpha=0.003$ & \cellcolor[gray]{0.8}$1-\beta=0.60$ \\ \hhline{|~|-|-|-|-|-|-|-|-|-|}
& 8 & \cellcolor[gray]{0.8}$\alpha=0.19$ & \cellcolor[gray]{0.8}$1-\beta=\times$& &\cellcolor[gray]{0.8}$\alpha=0.04$ &\cellcolor[gray]{0.8}$1-\beta=\times$& &$\alpha=0.005$ & \cellcolor[gray]{0.8}$1-\beta=0.73$\\ \hhline{|~|-|-|-|-|-|-|-|-|-|}
& 9 & \cellcolor[gray]{0.8}$\alpha=0.23$ & \cellcolor[gray]{0.8}$1-\beta=\times$& &\cellcolor[gray]{0.8}$\alpha=0.05$ &\cellcolor[gray]{0.8}$1-\beta=\times$& &$\alpha=0.008$ & $1-\beta=0.83$ \\ \hhline{|~|-|-|-|-|-|-|-|-|-|} 
& 10 & \cellcolor[gray]{0.8}$\alpha=0.26$ & \cellcolor[gray]{0.8}$1-\beta=\times$& &\cellcolor[gray]{0.8}$\alpha=0.07$& \cellcolor[gray]{0.8}$1-\beta=\times$& &$\alpha=0.013$ & $1-\beta=0.89$ \\ \hhline{|~|-|-|-|-|-|-|-|-|-|}
& 11 & \cellcolor[gray]{0.8}$\alpha=0.30$ & \cellcolor[gray]{0.8}$1-\beta=\times$& &\cellcolor[gray]{0.8}$\alpha=0.09$ & \cellcolor[gray]{0.8}$1-\beta=\times$& & $\alpha=0.019$ & $1-\beta=0.93$ \\ \hhline{|~|-|-|-|-|-|-|-|-|-|}
& 12 & \cellcolor[gray]{0.8}$\alpha=0.34$ & \cellcolor[gray]{0.8}$1-\beta=\times$& &\cellcolor[gray]{0.8}$\alpha=0.11$ & \cellcolor[gray]{0.8}$1-\beta=\times$& & \cellcolor[gray]{0.8}$\alpha=0.0256$ &\cellcolor[gray]{0.8}$1-\beta=\times$  \\ \hline \hline
\end{tabular}
\end{threeparttable}
\end{minipage}
}
\label{tab;example_0.1_0.55_abeta}
\end{table}

Finally, we set the lower limit $l_k$ for futility stopping.
Based on deterministic curtailment, it is determined as futility stopping when the number of responses cannot reach $u$ even though all of the subjects who are not known the response status make responses.
Thus, $l_k$ is set as follow: $l_K=u-1$ and $l_k=l_{k+1}-1(k=1,2,...,K-1)$. 
If $S_k\leq l_k$ in $k$th stage, stop the monitoring and reject $\mathcal{H}_1$. If $l_k<S_k<u$ in $k$th stage, continue the trial.

\end{spacing}

\section{Simulation 1: Design comparison in sample size}
\begin{spacing}{2.0}
\label{simulation_study_sample}
We compared the proposed design with fixed design, Lan-DeMets' $\alpha$-spending function with O’Brien-Fleming type (OF), and Simon's two-stage design (\citealp{gordon1983discrete,simon1989optimal}). To compare with the proposed design, we conducted an interim analysis of OF in each subject. We set 2 types of OF design, called OF-Wald and OF-Score, as follows: 
\begin{equation*}
N=\frac{(z_\alpha+z_\beta)^2p_1(1-p_1)}{(p_1-p_0)^2},
\end{equation*}
and
\begin{equation*}
N=\left(\frac{z_\alpha\sqrt{p_0(1-p_0)}+z_\beta\sqrt{p_1(1-p_1)}}{p_1-p_0}\right)^2.
\end{equation*}
where $N$ is the sample size and $z_q$ is the $100(1-q)$th percentile of the normal distribution. 

For instance, Table \ref{tab;example_0.1_0.35_threshold_example} shows threshold for each design with $(p_0,p_1)=(0.1,0.35)$ when nominal $\alpha$ error rate and nominal power are 0.025 and 0.8, respectively.
The null hypothesis $\mathcal{H}_0$ is rejected if the number of responses is 6 or higher by the time the maximum sample size of 22 is reached in the proposed design. By contrast, $\mathcal{H}_1$ is rejected if none of the subjects respond in 17 subjects. 

\begin{table}[htbp]
\centering
\caption{Thresholds for each design in $(p_0, p_1)=(0.1,0.35)$.}
\begin{threeparttable}
\begin{tabular}{c|cccccccccccccccc} \hline \hline
$k$ & 4 & 5 & 6 & 7 & 8 & 9 & 10 & \ldots & 17 & \ldots & 22 & \ldots & 25 & \ldots&30& 31 \\ \hline \hline
{\bf Proposed} & \\ \hline 
$u$ &-&-&6&6&6&6&6&\ldots&6&\ldots&6\\
$l_k$ &-&-&-&-&-&-&-&\ldots& 0 & \ldots & 5 \\ \hline
{\bf Fixed} &-&-&-&-&-&-&-&\ldots&-&\ldots&-&\ldots&7 \\ \hline
{\bf OF} & \\ \hline
Wald &-&-&6&6&7&7&7&\ldots&8&\ldots&8&\ldots&9&\ldots&9&9 \\
Score &4&4&4&4&4&4&4&\ldots&5& \\  \hline
{\bf Simon} & \\ \hline
Minimax &-&-&-&-&-&-&1&\ldots&-&\ldots&5 \\
Optimal &-&-&-&-&1&-&-&\ldots&-&\ldots&-&\ldots&-&\ldots&6\\ \hline \hline
\end{tabular}
\end{threeparttable}
\label{tab;example_0.1_0.35_threshold_example}
\end{table}

We determined the threshold in OF-Wald design. We found the minimum number of response satisfied with $|Z| > Z_{OF}$ using the test statistics $Z_{OF}$ calculated from $\alpha$-spending function method with $(p_0,p_1)=(0.1,0.35)$ and the Wald test statistics $Z$. These values are shown in Table \ref{tab;example_0.1_0.35_threshold_example}.
The thresholds are calculated for the Score test statistics in the same way. Here, in OF-Wald design, we can get $\hat{p}=1$ when the values of sample size and response are equal. Then, we used $Z$ with $\tilde{p}$ instead of $\hat{p}$ as in (\ref{agresti}) because $\sqrt{\frac{\hat{p}(1-\hat{p})}{k}}=0$ with Wald method \citep{agresti2018introduction},
\begin{equation*}
\label{agresti}
Z=\frac{\sqrt{k}(\tilde{p}-p_0)}{\sqrt{\tilde{p}(1-\tilde{p})}},\ {\rm where}\ \tilde{p}=\hat{p}\left(\frac{k}{k+z^2_\alpha}\right)+\frac{1}{2}\left(\frac{k}{k+z^2_\alpha}\right)=\frac{\hat{p}+\frac{z_\alpha^2}{2k}}{1+\frac{z_\alpha^2}{k}}.
\end{equation*}

The performance measures are maximum sample size, average sample number and power. The replication number was 100,000. We set the true response rate $p$ from 0.05 to 0.6 by 0.05, the null response rate $p_0$ from 0.1 to 0.3 by 0.1, and the expected response rate $p_1$ from 0.25 to 0.5 by 0.05.
The overall $\alpha$ error rate and overall power were 0.025 and 0.8.
Maximum sample sizes in OF design were calculated using R package gsDesign (\citealp{gsDesign2021}). Similarly, sample sizes in Simon’s 2-stage design were calculated using R package clinfun (\citealp{clinfun2018}).

Table \ref{tab;mss_table_summary} shows the maximum sample size.
OF-Score design observed a significant inflated $\alpha$ error rate to the nominal level in some scenarios. Thus, we will not discuss the sample sizes of OF-Score design.
The maximum sample size of the proposed design are not only smaller than that of most of the sequential tests but equal to or smaller than that of the fixed design (Fixed). 
Some values in Simon's minimax design are slightly smaller than the proposed design for some pairs $(p_0, p_1)$, but the value in Simon's minimax design approaches the value in proposed design as $p_1$ increased under the same value in $p_0$.

\begin{table}[htbp]
\centering
\caption{Maximum sample size for each design.}
\begin{threeparttable}
\begin{tabular}{c||cccccc|cccc|ccc} \hline \hline
$p_0$ & 0.1 &&&&&& 0.2 &&&& 0.3 & \\
$p_1$ & 0.25 & 0.3 & 0.35 & 0.4 &0.45 &0.5 &0.35 &0.4 &0.45 &0.5 &0.45 &0.5  \\ \hline \hline
{\bf Proposed} & 49 & 29 & 22 & 16 & 11 & 10 & 72 & 41 & 26 & 19 & 83 & 47 \\ \hline
{\bf Fixed} & 53 & 33 & 25 & 19 & 14 & 10 & 78 & 44 & 31 & 24 & 88 & 54 \\ \hline
{\bf OF} &&&&&&&&&& \\ \hline
Wald & 70 & 45 & 31 & 23 & 17 & 14 & 85 & 51 & 34 & 24 & 92 & 52 \\ 
Score & 44 & 26 & 17 & 13 & 10 & 8 & 67 & 38 & 26 & 18 & 82 & 47 \\ \hline \hline
{\bf Simon} &&&&&&&&&& \\ \hline
Minimax & 49 & 29 & 22 & 16 & 11 & 10 & 69 & 41 & 26 & 19 & 81 & 47 \\
Optimal & 58 & 38 & 30 & 18 & 12 & 11 & 83 & 55 & 35 & 23 & 100 & 65 \\ \hline \hline
\end{tabular}
\end{threeparttable}
\label{tab;mss_table_summary}
\end{table}

\begin{figure}[htbp]
\begin{center}
\includegraphics[scale=0.3]{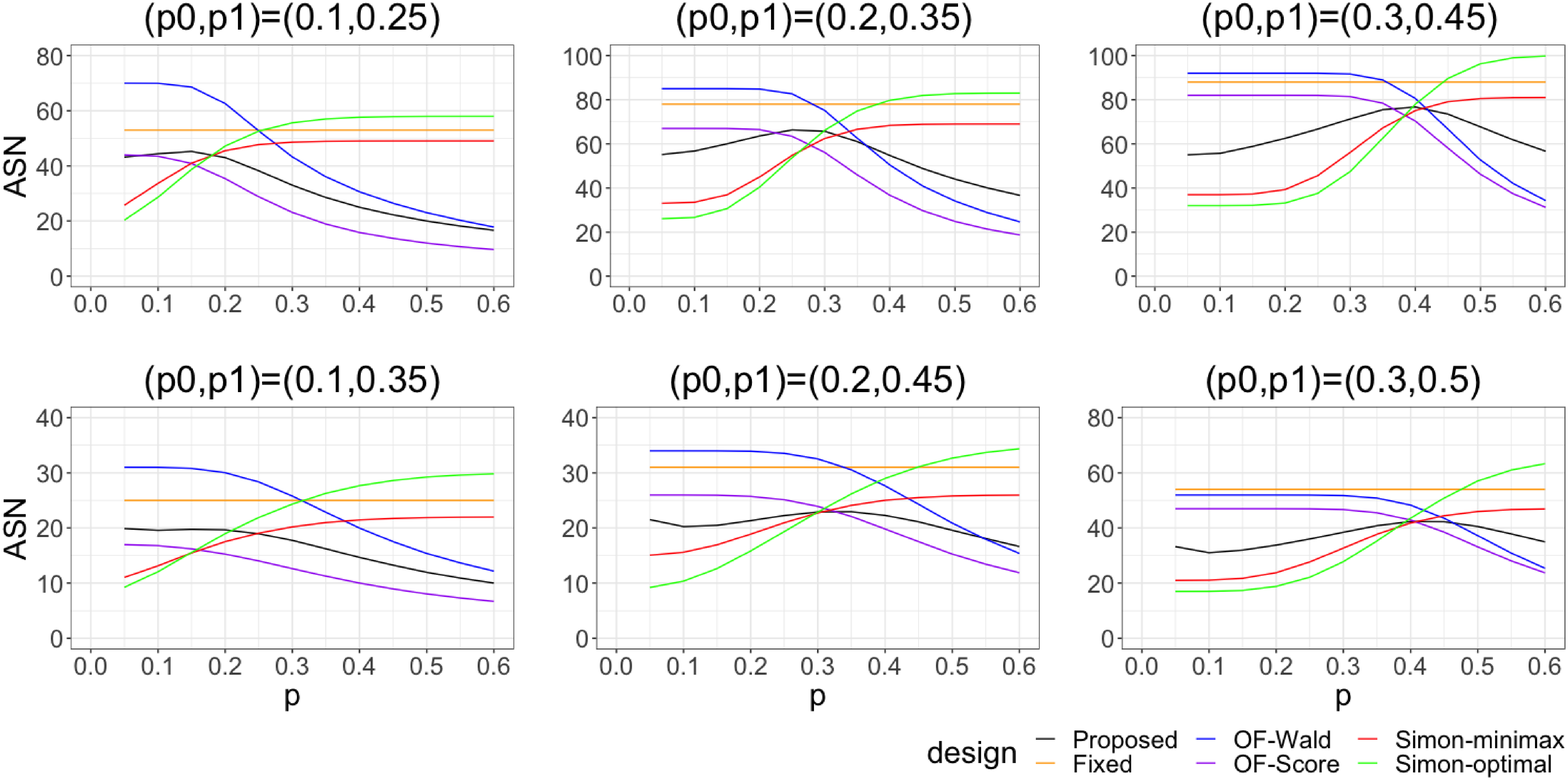}
\end{center}
\caption{Average sample number against the true value of $p$ for each design in some hypothesis}
\label{asn_plot}
\end{figure}

As a result of simulations, the actual power is 0.8 or higher when $p\geq p_1$ for all tests. Figure \ref{asn_plot} shows the values of the average sample number for each test. The average sample number for the proposed design is getting smaller as the $p$ gets larger in $p\geq p_1$.
Furthermore, it is also small as the $p$ closing to $p_0$ due to deterministic curtailment.
The average sample number of the proposed design are only larger than that of both Simon's designs under small $p$ and that of OF-Wald design under $p_0\geq 0.2$ and large $p$.
\end{spacing}

\section{Estimation}
\label{estimation}
\begin{spacing}{2.0}
A point estimate of the response rate and its confidence interval, which is calculated at the termination of the trial, are useful to design the future phase III trials and other phase II trials. Generally, the naive point estimator which is not taken into account of an interim monitoring may be biased(\citealp{guo2005simple}). In this section, we will discuss bias-adjusted point estimation and the construction of confidence interval.

Let $(M,S)$ be the random variables that terminate the trial at the $M$th interim analysis with the number of response $S$. Due to sequential monitoring in our settings, the number of enrolled subjects is also $M$. \citet{jung2004estimation} derive the probability mass function of the random vector $(M,S)$ for observed values $(m,s)$ as
\begin{equation*}
f(m,s|p)=c_{m,s}p^s(1-p)^{m-s}
\end{equation*}
with support $\mathscr{S}=\bigcup_{m=1}^K\mathscr{S}_m$, where
\begin{equation*}
\mathscr{S}_m=\{(m,s)|\ l_{m-1}+1\leq s \leq l_m\ {\rm or}\ s=u\},
\end{equation*}
and $c_{m,s}$ represents the sum of number of paths which the monitoring stopped at the $m$th interim analysis with the number of response $s$ . $c_{m,s}$ is defined as 
\begin{equation*}
c_{m,s}=\sum_{x_1}\sum_{x_2}...\sum_{x_m}\binom{1}{x_1}\binom{2}{x_2}...\binom{m}{x_m}
\end{equation*}
with the summations over the set
\begin{equation*}
\begin{split}
\mathscr{R}(m,s)=\{&(x_1,...,x_m):x_1+...+x_m=s,\ l_k+1<x_1+...+x_k<u-1\\ &{\rm for}  \ k=1,...,m-1\}.
\end{split}
\end{equation*}

The naive maximum likelihood estimator $\hat{\pi}_n=\frac{s}{m}$ is biased when ignoring a sequential monitoring process, and \citet{guo2005simple} developed bias-adjusted estimator $\hat{\pi}_g$ as the numerical solution of
\begin{equation*}
\hat{\pi}_g=\hat{\pi}_n-B(\hat{\pi}_n),
\end{equation*}
where the bias is given by
\begin{equation*}
B(p)=E(\hat{p}|p)-p=\sum_{(m,s)\in\mathscr{S}}\hat{p}f(m,s|p)-p.
\end{equation*}
This estimator has advantages that reduce computational complexity compared with \citet*{chang1989bias}'s estimator without the loss of accuracy. 
However, it is not always feasible or desirable to use this estimator due to the bias-variance trade-off. In addition, the computation of estimators applying the Rao-Blackwell theorem may not be feasible when estimating in a flexible design such as an adaptive design (\citealp{robertson2021point}). For these reasons, a median unbiased estimator is also proposed. For the binomial outcomes, \citet{jovic2010exact} proposed using two $p$-value functions $P(p),Q(p)$ defined by  
\begin{equation*}
P(p)=
\begin{cases}
Pr(S\geq s|p) \ \ \ \ \ \  & {\rm if}\ s\geq u   \\
1-Pr(S< s|p) \ \ \ \ \ \ & {\rm if}\ s\leq l_k
\end{cases}
\end{equation*}
and
\begin{equation*}
Q(p)=
\begin{cases}
Pr(S> s|p) \ \ \ \ \ \  & {\rm if}\ s\geq u   \\
1-Pr(S\leq s|p) \ \ \ \ \ \ & {\rm if}\ s\leq l_k.
\end{cases}
\end{equation*}
The other $p$-value function named ``stage-wise ordering'' is defined as
\begin{equation*}
P(p)=
\begin{cases}
Pr((M,S)\succeq (m,s)|p) \ \ \ \ \ \  & {\rm if}\ s\geq u   \\
1-Pr((M,S)\prec (m,s)|p) \ \ \ \ \ \ & {\rm if}\ s\leq l_k
\end{cases}
\end{equation*}
and
\begin{equation*}
Q(p)=
\begin{cases}
Pr((M,S)\succ (m,s)|p) \ \ \ \ \ \  & {\rm if}\ s\geq u   \\
1-Pr((M,S)\preceq (m,s)|p) \ \ \ \ \ \ & {\rm if}\ s\leq l_k.
\end{cases}
\end{equation*}
Here, $A \succeq B$ means that $A$ is later than or at the same stage as $B$ in stage-wise ordering. \citet{jennison1983confidence,jennison1999group} recommended this ordering for sequential clinical trials. For instance, the proposed method under $(p_0,p_1)=(0.1,0.35)$ gives the stopping boundary shown in Table \ref{tab;example_0.1_0.35_threshold_example}. So the sequential process realize the below pair and its stage-wise ordering are represented as
\begin{equation*}
(17,0) \preceq (18,1) \preceq (19,2) \preceq (20,3) \preceq (21,4) \preceq (22,5) \preceq (22,6) \preceq (21,6) \preceq (20,6) \preceq ... \preceq (6,6).
\end{equation*}
An median unbiased estimate is denoted by $\pi_h$, where $\pi_h=\frac{1}{2}(\pi_h^-+\pi_h^+)$ with $P(\pi_h^-)=0.5$ and $Q(\pi_h^+)=0.5$. The calculation methods of $\pi_h$ were studied by \citet{jovic2010exact}, they pointed out  substantial overestimation of $p$ by normal approximation when $p$ is not close to $p_0$. The exact method proposed by their study is close to the naive estimate in fixed design and keeps the conservative coverage probability.

Next, we take up the topic of constructing confidence interval. Even if one-sided test is used, two-sided $(1-2\alpha)$ confidence interval is often reported. According to \citet{clopper1934use}, the confidence interval $(p_L,p_U)=(p_L(S),p_U(S))$ are constructed as the following:
\begin{equation*}
\begin{split}
Pr[S\leq s|p=p_U(s)]=\sum_{i=0}^{S} \binom{m}{i} p_U^i (1-p_U)^{k-i}=\alpha \\
Pr[S\geq s|p=p_L(s)]=\sum_{i=S}^{m} \binom{m}{i} p_L^i (1-p_L)^{k-i}=\alpha,
\end{split}
\end{equation*}
where they are defined by $p_L(0)=0$, $p_U(k)=1$. Another approach given by \citet{jennison1983confidence} is to use Clopper-Pearson's method of an exact confidence interval using the appropriate distribution of $(M,S)$. This defines the interval $(p_L(s),p_U(s))$ as
\begin{equation*}
\begin{split}
Pr[(M,S)\succeq (m,s)|p=p_U(s)]=\alpha \\
Pr[(M,S)\preceq (m,s)|p=p_L(s)]=\alpha.
\end{split}
\end{equation*}
To make the coverage probability close to the nominal level, it has been proposed to use the mid-$p$ approach. Applying this approach to Clopper-Pearson's interval (\citealp{berry1995mid}) yields the following equation: 
\begin{equation*}
\begin{split}
Pr[S<s|p=p_U(s)]+\frac{1}{2}Pr[S=s|p=p_U(s)]=\alpha \\
Pr[S>s|p=p_L(s)]+\frac{1}{2}Pr[S=s|p=p_L(s)]=\alpha.
\label{mid-p_CP}
\end{split}
\end{equation*}
According to \citet{porcher2012inference}, we can derive \citet{jennison1983confidence}'s confidence interval with mid-$p$ approach as follows:
\begin{equation*}
\begin{split}
\label{mid-p_JT}
Pr[(M,S)\succ (m,s)|p=p_U(s)]+\frac{1}{2}Pr[(M,S)=(m,s)|p=p_U(s)]=\alpha \\
Pr[(M,S)\prec (m,s)|p=p_L(s)]+\frac{1}{2}Pr[(M,S)=(m,s)|p=p_L(s)]=\alpha.
\end{split}
\end{equation*}
The upper and lower probabilities $(p_L,p_U)$ are calculated separately in the above methods.

\citet{duffy1987confidence} proposed the sequential calculation of acceptance regions. First, we calculate $f(m,s|p_i)$ from $p_i$ separated small enough within $[0,1]$ and the aligned pairs $(m,s)$ based on two linear ordering.
In the following simulation experiments, we calculated on a grid of every 0.0001.
Next, we constitute a minimum cardinality acceptance region $A(p_i)=[L(p_i),...,U(p_i)]$ satisfying the conditions (i)-(iii) written in this paper.
Last, we determine $p_L$ and $p_U$ be the smallest and largest $p_i$ for which $A(p_i)$ contains the observed value $(m,s)$, respectively.
\end{spacing}

\section{Simulation 2: Point estimate and confidence interval}
\label{simulation_study_est}
\begin{spacing}{2.0}

We evaluated the comparison for point estimates and confidence intervals introduced in section \ref{estimation}. In point estimates, we took up a maximum likelihood estimator $\hat{\pi}_n$(Naive), a bias-adjusted estimator $\hat{\pi}_g$ (Bias-adjusted), and a median unbiased estimator using stage-wise ordering in the composition of two $p$-value functions $\hat{\pi}_h$ (MUE). The step size for computing the values $\pi_h^-, \pi_h^+$ was set to 0.0001. Bias and root mean square error (RMSE) were evaluated for these estimators. In confidence intervals, we also took up Clopper and Pearson(CP) interval, Jennison and Turnbull(JT) interval, CP's interval using the mid-$p$ approach (mid$p$-CP), JT's tail interval using mid-$p$ approach (mid$p$-JT), and multistage Duffy and Santner interval (DufSat). The performance measures were coverage probability and the average of length in confidence intervals. The replication number was 10,000, and the scenarios for $p_0$, $p_1$, and $p$ were the same as Section \ref{simulation_study_sample}).
\begin{figure}[htbp]
\begin{center}
\includegraphics[scale=0.28]{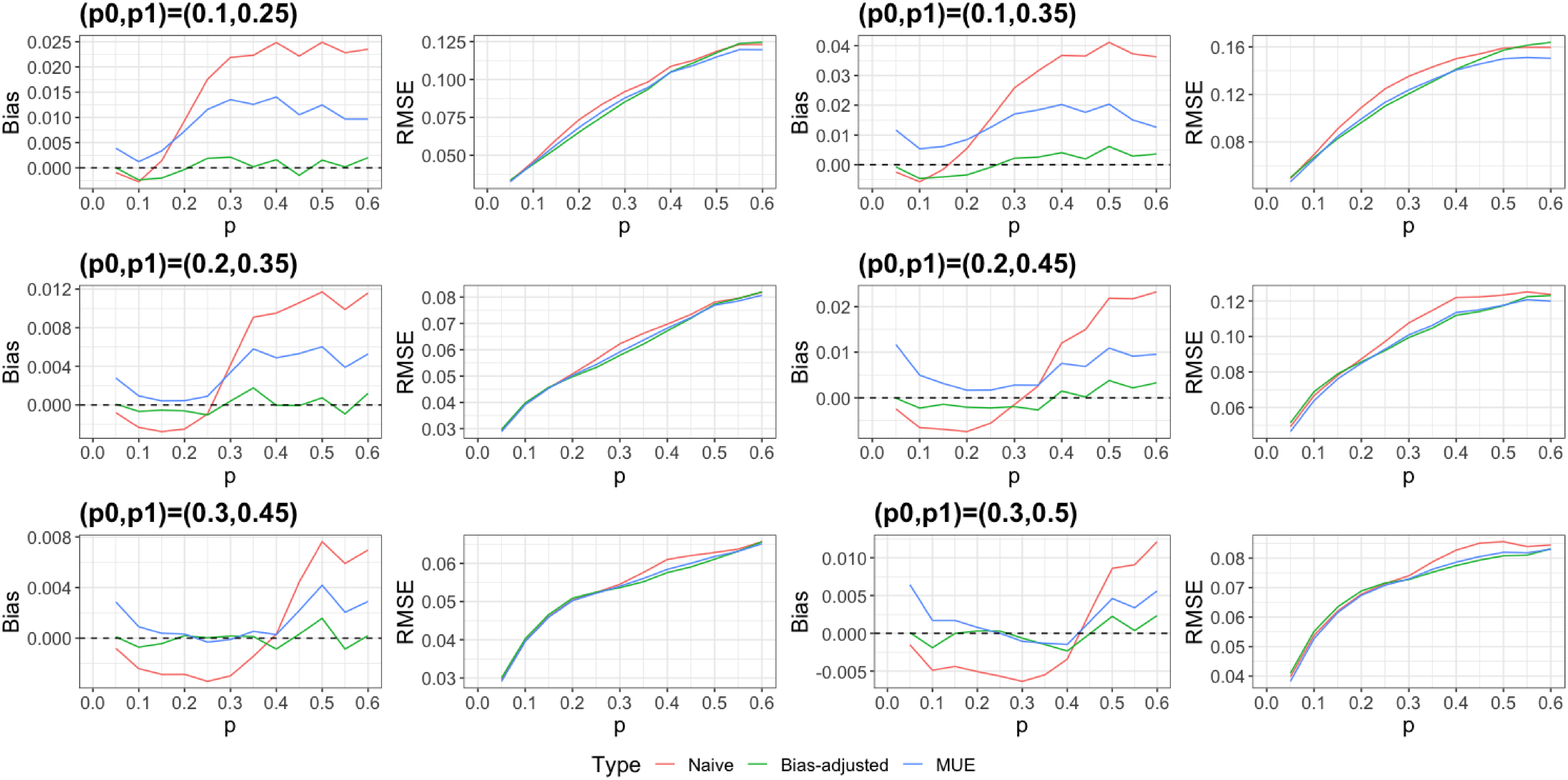}
\end{center}
\caption{Bias and RMSE in the point estimation.}
\label{bias_rmse_plot}
\end{figure}

\begin{figure}[htbp]
\begin{center}
\includegraphics[scale=0.28]{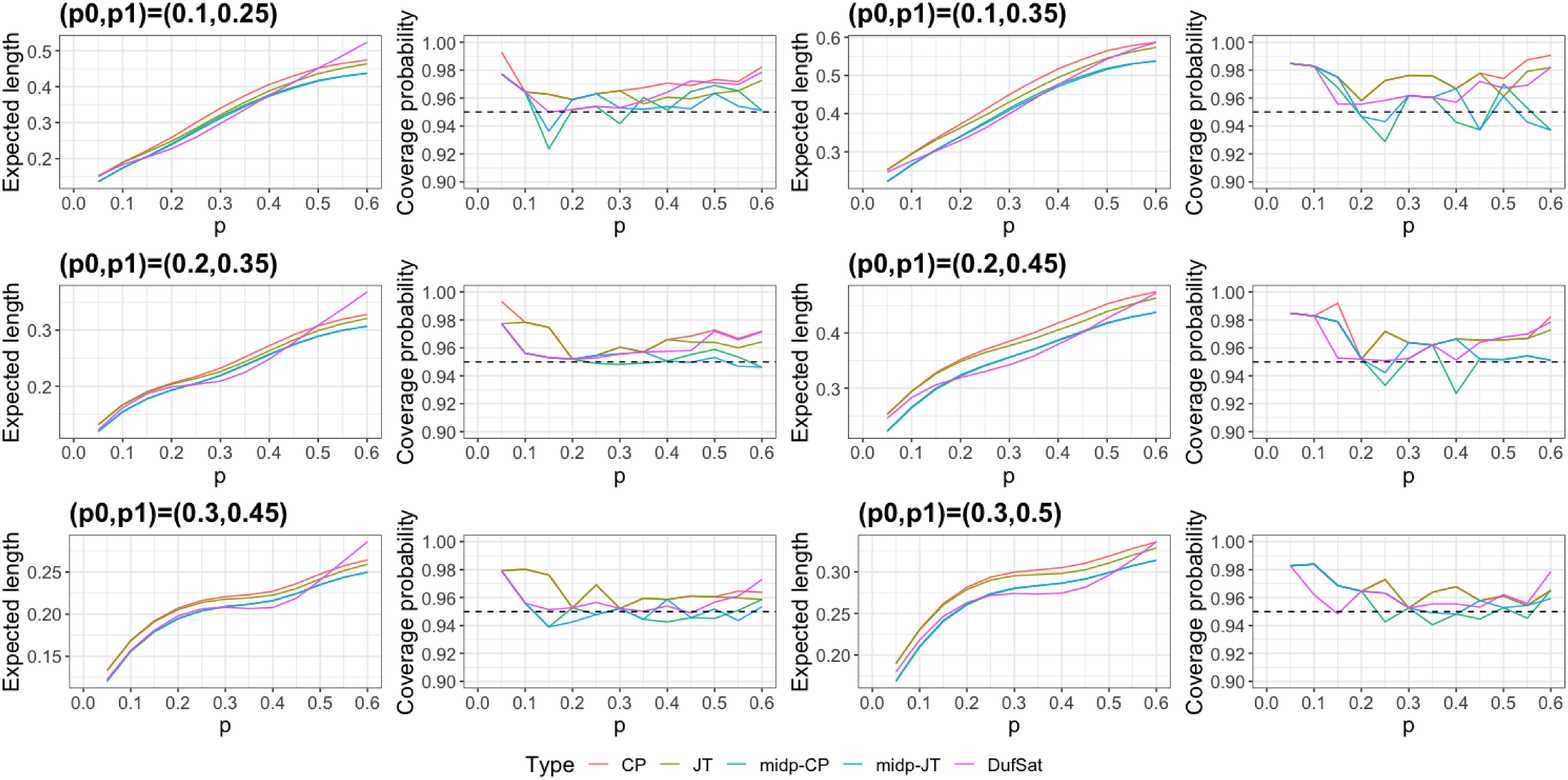}
\end{center}
\caption{Expexted length and coverage probability in the 95\% confidence interval.}
\label{exp_cov_plot}
\end{figure}
The bias and RMSE are shown in Figure \ref{bias_rmse_plot}.
Both the absolute bias and RMSE of bias-adjusted estimator tended to be smallest than that of the naive estimator and MUE. Figure \ref{exp_cov_plot} illustrates the average length and coverage probability in the 95\% confidence intervals.
The DufSat, CP, and JT exceeded the nominal level of 95\% in all scenarios of true response rate $p$. The midp-CP and midp-JT had at least about 92\% coverage probability.
DufSat had the shortest intervals under $p$ being near or less than the expected response rate $p_1$. For $p$ is exceed $p_1$, the length of the mid-p JT interval was shortest than that of other methods.
\end{spacing}

\section{Discussion}
\label{discussion}
\begin{spacing}{2.0}
We developed the single-arm exact sequential design for a binary endpoint. This design has the fixed threshold for the efficacy stopping independent of the number of interim monitoring, that leads to avoid paying close attention to the number of enrollments.
The sample size, actual $\alpha$ error rate and power can be calculated using cumulative negative binomial distribution function. The design also enables the stopping for futility by deterministic curtailment. Furthermore, we explored the bias-adjustment estimator and construction of confidence interval for the proposed monitoring method. 

Remarkably, the maximum sample size of the proposed sequential design is smaller than that of the fixed design, though a maximum sample size of a sequential design generally tends to be larger than that of a fixed design because of adjustment to repeat statistical tests. In a small sample, the actual power increases or decreases with increasing sample size because there is discreteness in the threshold number of responses to determine significance. To allow for an overshoot in enrollment, it determines the sample size that, if larger than that size, always achieves a higher power than a nominal level.
There may be a certain sample size with enough power less than a fixed sample size.
Since the power of our method is a weighted sum of each sample size, we consider that the smaller sample size than the conservatively calculated sample size in a fixed design would have maintained sufficient power.
The average sample number of the proposed design is similar to that of Simon's two-stage design under less or no treatment effect and is also similar to that of the spending function approach under the presence of treatment effect. We think such a versatile feature encourages us to adopt it for single-arm binary trials.

Regarding the point estimator, we think the bias-adjusted estimator $\hat{\pi}_g$ is most suitable than the naive estimator and MUE because of the smallest bias and RMSE. 
Any methods of constructing confidence interval had a reasonable coverage probability in the simulation experiments. DufSat method had the shortest intervals under true response rate being near or less than expected response rate. In the other scenarios, the length of the mid-p JT interval was shortest than that of other methods. When the response rate is greater than expected, a wide confidence interval will be easy to accept in interpretation. Thus we suggested that the bias-adjusted point estimator and DufSat confidence interval are recommended in the estimation for response rate.

We evaluated limited situations in which proposed design has deterministic curtailment. If replacing deterministic curtailment with stochastic curtailment, we expect to getting a smaller sample size in stopping for futility. In addition, it would be useful to consider the uniformly minimum variance unbiased estimator due to bias and variance trade-off. We have already derived this estimator in the framework of proposed design based on \citet{jung2004estimation}, there may be better modification to reduce the bias. 
\end{spacing}

\bibliographystyle{biom}
\bibliography{ENAR_hokudai_inao_Bibliography}

\end{document}